\documentclass[journal,10pt]{IEEEtran}

\usepackage{cite}
\usepackage{graphicx}
\usepackage{amsmath}
\usepackage{array}
\usepackage{mdwmath}
\usepackage{amssymb}
\usepackage{mdwtab}
\usepackage{stfloats}
\usepackage[tight,footnotesize]{subfigure}
\usepackage{amsmath,amsthm}
\usepackage{threeparttable}
\usepackage{color}
\usepackage{url}
\usepackage{algpseudocode}
\usepackage{algorithm}
\usepackage{multicol}
\usepackage{amssymb}
\usepackage{epstopdf}

\algrenewcommand{\algorithmicrequire}{\textbf{Input:}}
\algrenewcommand{\algorithmicensure}{\textbf{Output:}}

\begin{document}

\title{\LARGE Movable Antenna Empowered Downlink NOMA Systems: Power Allocation and Antenna Position Optimization}

\author{Yufeng Zhou, Wen Chen, \IEEEmembership{Senior Member, IEEE}, Qingqing Wu, \IEEEmembership{Senior Member, IEEE}, Xusheng Zhu, Nan Cheng, \IEEEmembership{Senior Member, IEEE}% stops a space
\thanks{Y. Zhou, W. Chen, Q. Wu, and X. Zhu are with the Department of Electronic Engineering, Shanghai Jiao Tong University, Shanghai 200240, China (e-mail: ereaked@sjtu.edu.cn; wenchen@sjtu.edu.cn; qingqingwu@sjtu.edu.cn; xushengzhu@sjtu.edu.cn).
	
N. Cheng is with the School of Telecommunications Engineering, Xidian University, Xi'an 710071, China (e-mail: nancheng@xidian.edu.cn).}% <-this % stops a space
%\thanks{}% <-this % stops a space
}

\markboth{}
{}
\maketitle

\begin{abstract}
This paper investigates a novel communication paradigm employing movable antennas (MAs) within a multiple-input single-output (MISO) non-orthogonal multiple access (NOMA) downlink framework, where users are equipped with MAs. Initially, leveraging the far-field response, we delineate the channel characteristics concerning both the power allocation coefficient and positions of MAs. Subsequently, we endeavor to maximize the channel capacity by jointly optimizing power allocation and antenna positions. To tackle the resultant non-concave problem, we propose an alternating optimization (AO) scheme underpinned by successive convex approximation (SCA) to converge towards a stationary point. Through numerical simulations, our findings substantiate the superiority of the MA-assisted NOMA system over both orthogonal multiple access (OMA) and conventional NOMA configurations in terms of average sum rate and outage probability.
\end{abstract}


\section{Introduction}
Multiple-input multiple-output (MIMO) stands as a cornerstone technology in the fifth-generation (5G) wireless standard, enhancing the efficiency of wireless communication through the exploitation of innovative degrees of freedom (DoFs) within the spatial domain. Numerous researchers leverage MIMO technology, integrating it with other promising techniques, yielding notable outcomes \cite{Zhang2024Int, Zhu2023Per, Wang2017Joint}. Nevertheless, the antenna arrangement in conventional MIMO systems remains fixed, lacking adaptability to accommodate abrupt channel fading.

In response to the limitations inherent in conventional antenna configurations, a pioneering antenna arrangement, namely the movable antenna (MA) was introduced in \cite{Zhu2023Mod}. This innovative design permits the antenna to maneuver within a two-dimensional (2D) region, thereby tacking sudden channel fading by dynamically adjusting the signal phases across various propagation paths upon reaching the receivers. In \cite{Qin2024Antenna}, researchers investigated a multi-user downlink communication system featuring a BS equipped with fixed position antennas (FPAs) and users possessing a single MA. It is illustrated that this configuration could achieve lower power consumption under equivalent outage probability conditions through the joint optimization of beam-forming at the BS and antenna positions adjusting at the users. The authors in \cite{Cheng2024Sum} considered a multi-user downlink system wherein the BS is outfitted with MAs, while the users possess a single FPA. Furthermore, the researchers in \cite{Mei2024Mov} convert the search for optimal MA continuous positions into a discrete optimization problem, seeking the best point among those sampled from the transmission region.

On the other hand, non-orthogonal multiple access (NOMA) stands as a pivotal multiple access technique in bolstering the forthcoming Internet-of-Things (IoTs) and mobile internet landscapes \cite{Liu2017Non, Liu2022Evo, Huang2023Coe}. The fundamental premise of NOMA resides in enabling disparate users to share the same communication resource (i.e., time, frequency, code), while distinguishing their respective signals through varying power allocation. Researchers demonstrated that systems employing NOMA can attain superior spectrum and power efficiency than that of adopting conventional orthogonal multiple access (OMA) as indicated in \cite{Ding2014On, Liu2016Coo}.

Inspired by the aforementioned description, we consider the integration of NOMA and MA. To date, research on MAs is abundant \cite{Zhu2023Movable, Wang2024Movable, Gao2024Joint}. For instance, the authors of \cite{Zhu2023Movable} investigate combining beamforming with movable antennas in BSs, achieving full array gain in desired directions and null steering in undesired ones. However, the study of combining NOMA with MAs remains limited; only the researchers of \cite{Li2024Sum} have explored this combination in uplink communication systems. We investigate the MA-assisted NOMA downlink communication system, and then we formulate the problem of maximizing the channel capacity by joint optimization of power allocation and antenna positions. Given the inherent non-concave nature of this problem, rendering it challenging to solve directly, we propose an alternating optimization (AO) algorithm capitalizing on successive convex approximation (SCA) to iteratively converge towards a stationary point.

\section{System Model and Problem Formulation}
As depicted in Fig. 1, we consider a multiple-input single-output (MISO) downlink communication architecture incorporating NOMA and MA technologies. Within this framework, a BS featuring a constellation of $N$ FPAs caters to two distinct users each equipped with a single MA. We designate the first user, termed as the user1, as proximal to the BS relative to the second user, denoted as the user2.

The position of the MA corresponding to the $i$-th user can be represented as $\mathbf{r}_i = \left[x^r_i, y^r_i\right]^T \in \mathcal{D}_i, i={1,2}$, by establishing via 2D local coordinate system, where $\mathcal{D}_i$ is the 2D square region for MA of the $i$-th user moving with the identical size of $A \times A$. Similarly, The position of the $k$-th antenna at BS can be described as $\mathbf{t}_k = \left[x^t_k, y^t_k\right]^T,1 \leq k \leq N$. Hence, the received baseband signal of the $i$-th user can be expressed as follows.
\begin{equation}
	y_i = h_i(\mathbf{r}_i)\cdot \sqrt{P_0} \cdot (\sqrt{\alpha} \cdot s_1 + \sqrt{1-\alpha} \cdot s_2) + n_i,
\end{equation}
where $P_0$ is the transmitting power of each antenna at BS and the total power of BS is $P \triangleq N \cdot P_0$; $h_i(\mathbf{r}_i) \in \mathbb{C}^{1\times 1}$ is the channel response coefficient (CRC) between the BS and user$i$, $i={1,2}$; $s_1, s_2 \in \mathbb{C}^{1\times1}$ denote signals sent to user1 and user2 respectively with normalized power, i.e., $\mathbb{E}(s_i^*s_i)=1, i={1,2}$; $\alpha$ is the power allocation coefficient for the user1; and $n_i$ represents the addictive white Gaussian noise (AWGN) in received signal of user$i$ with average power $\sigma^2$, i.e., $\mathbf{n} = \left[n_1, n_2\right]^T \sim \mathcal{CN} \left(\mathbf{0}, \sigma^2\mathbf{I_2}\right)$.

\begin{figure}
	\centering
	\includegraphics[width=1\linewidth]{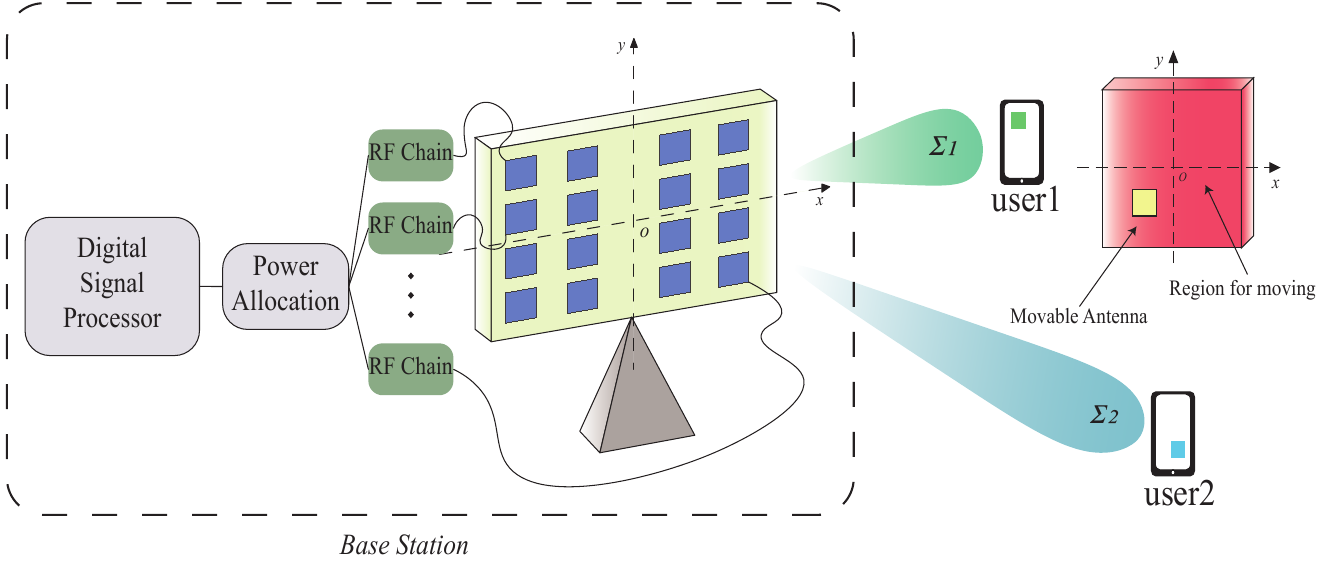}
	\caption{MA-assisted NOMA downlink communication system.}
\end{figure}

In this system, we focus on a field response channel\cite{Zhu2023Mod}. The number of transmit paths from the BS origin $\mathbf{o}^t = [0, 0]^T$ to the origin of the $i$-th user $\mathbf{o}^r_i = [0, 0]^T$ is $L_i^t$; similarly, the number of receive paths is $L_i^r$. Therefore, the path response matrix (PRM) between the BS and the $i$-th user is defined as $\mathbf{\Sigma}_i \in \mathbb{C}^{L^r_i \times L^t_i}$, where $\mathbf{\Sigma}_i\left[n,m\right]$ represents the CRC between the $m$-th transmit path and the $n$-th receive path from the BS origin to that of the user $i$. According to the far field response model, the movements of MA are not anticipated to alter the angle-of-departure (AoD), the angle-of-arrival (AoA), or the amplitude of the CRC for each channel path extending from the BS to users. For the $i$-th user, the discrepancy in signal propagation for the $m$-th transmit path between the position of $k$-th antenna $\mathbf{t}_k = [x_k, y_k]^T$ and original point $\mathbf{0}^t = [0, 0]^T$ can be described as $\rho^t_{i,m}(\mathbf{t}_k)=-[x_k^t\cos\theta^t_{i,m}\sin\phi^t_{i,m}+y_k^t\sin\theta^t_{i,m}] $, where $\theta^t_{i,m}$ and $\phi^t_{i,m}$ represent the elevation and azimuth of the AoD for the $m$-th transmit path to user $i$ respectively. Consequently, its associated phase shift is denoted as $-j\frac{2\pi}{\lambda} \cdot \rho^t_{i, m}(\mathbf{t}_k)$, where $\lambda$ is the carrier wavelength. Subsequently, the transmit field response vector (FRV) of the $k$-th FPA at the BS can be derived as follows
\begin{equation}
	\mathbf{g}_i(\mathbf{t}_k) = [e^{-j\frac{2\pi}{\lambda}\rho^t_{i,1}(\mathbf{t}_k)}, e^{-j\frac{2\pi}{\lambda}\rho^t_{i,2}(\mathbf{t}_k)}, \cdots, e^{-j\frac{2\pi}{\lambda}\rho^t_{i,L_i^t}(\mathbf{t}_k)}]^T.
\end{equation}

Similarly, the receive FPV of the $i$-th user can be described as
\begin{equation}
	\mathbf{f}_i(\mathbf{r}_i) = [e^{-j\frac{2\pi}{\lambda}\rho^r_{i,1}(\mathbf{r}_i)}, e^{-j\frac{2\pi}{\lambda}\rho^r_{i,2}(\mathbf{r}_i)}, \cdots, e^{-j\frac{2\pi}{\lambda}\rho^r_{i,L_i^r}(\mathbf{r}_i)}]^T,
\end{equation}
where $\rho^r_{i,n}(\mathbf{r}_i) = x_i^r\cos\theta^r_{i,n}\sin\phi^r_{i,n} + y_i^r\sin\theta^r_{i,n}$ represents the signal propagation difference for $n$-th receive path between the position of MA $\mathbf{r}_i=[x_i^r,y_i^r]^T$ and original point $\mathbf{o}_i^r=[0,0]^T$ at user $i$. And $\theta^r_{i,n}, \phi^r_{i,n}$ denote the elevation and azimuth of the AoA for the $n$-th receive path to user $i$ separately.
Thus, the CRC between the BS and user $i$ can be obtained as
\begin{equation}
	h_i(\mathbf{r}_i) =\mathbf{f}_i(\mathbf{r}_i)^T\cdot \mathbf{\Sigma}_i \cdot \mathbf{G}_i \cdot \mathbf{1},
\end{equation}
where $\mathbf{G}_i = [\mathbf{g}_i(\mathbf{t}_1), \mathbf{g}_i(\mathbf{t}_2), \cdots, \mathbf{g}_i(\mathbf{t}_N)] \in \mathbb{C}^{L_i^t \times N}, i={1,2}$ is the transmit field response matrix (FRM) from the BS to the $i$-th user; $\mathbf{1} = [1, 1, \cdots, 1]^T \in \mathbb{R}^{N \times 1}$ represents the all-one vector.

In adherence to the fundamental concept of successive interference cancellation (SIC), the power allocation coefficient is lower for the user experiencing an robust channel condition, denoted as $\alpha_s \triangleq \min\{\alpha, 1-\alpha\}$. Thus, the signal-to-noise-plus-interference ratio (SINR) for the user encountering a robust channel condition is denoted as $\gamma_s$, while that of the user suffering an inferior channel condition is given as $\gamma_w$, as exhibited bellow.
\begin{align}
		\gamma_s &= \frac{P_0 \cdot \alpha_s \cdot \max{\{ \left|h_1(\mathbf{r}_1)\right|^2, \left|h_2(\mathbf{r}_2)\right|^2\}}}{\sigma^2}\\
		\gamma_w &= \frac{P_0 \cdot (1-\alpha_s) \cdot \min{\{ \left|h_1(\mathbf{r}_1)\right|^2, \left|h_2(\mathbf{r}_2)\right|^2 \}}}{P_0 \cdot \alpha_s \cdot \min{\{ \left|h_1(\mathbf{r}_1)\right|^2, \left|h_2(\mathbf{r}_2)\right|^2 \}} + \sigma^2}
\end{align}

The primary objective of this study is to maximize channel capacity through the joint optimization of both the power allocation coefficient and the position of MA for each user. We posit that successful decoding of the transmitted signal by users occurs when SINR surpasses a predetermined threshold $\gamma_0$. Drawing on the Shannon capacity theorem, this optimization problem can be formulated as
\begin{subequations}\label{eqn-1}
	\begin{align}
		\text{(P}&\text{0)}  \quad \max_{\{\alpha_s, \mathbf{r}_1, \mathbf{r}_2\}} \log_2 \left(1+\gamma_s\right)+ \log_2\left(1+\gamma_w\right), 		
		\\
		\text{s.}&\text{t.} \qquad\qquad\qquad\qquad
		 \gamma_s \geqslant\gamma_0,\\
		&\quad\quad\quad\quad
		\frac{\rho \cdot \left(1- \alpha_s\right) \cdot \left|h_i(\mathbf{r}_i)\right|^2}
		{\rho \cdot \alpha_s \cdot \left|h_i(\mathbf{r}_i)\right|^2+ 1}\geqslant \gamma_0, \enspace i={1,2},\\
		& \enspace\quad\quad\quad\quad\quad\quad\quad\enspace
		0\leqslant\alpha_s\leqslant1,\\
		& \quad\quad\quad\quad\quad\quad\quad\quad\quad
		\mathbf{r}_i\in \mathcal{D}_i, \quad\quad\quad\quad i={1,2},
	\end{align}
\end{subequations}
where $\rho=\frac{P_0}{\sigma^2}$ represents the ratio between the transmitting power of each antenna and the average noise power; constraints (7b), (7c) specify the minimal SINR necessary for accurate signal decoding by users; constraint (7e) mandates that the MAs of users remain within the designated square region. Due to the non-concave nature of the objective function and constraints (7b), (7c), the optimization problem is inherently non-concave, posing challenges for directly obtaining the optimal solution.

\section{Proposed Solution}
In this section, due to the complexity of optimizing $\mathbf{r}_1$, $\mathbf{r}_2$, and $\alpha_s$ simultaneously, we introduce an alternating optimization algorithm for addressing (P0), which involves decomposing it into two sub-problems, denoted as (P1) and (P2), respectively. The optimization process alternates between refining the power allocation coefficient and adjusting the positions of MAs while holding the other side constant. 
\subsection{Power Allocation Coefficient}
Initially, we fix the positions of MAs. Then, exploiting the monotonicity of the logarithmic function, we formulate sub-problem (P1) as follows
\begin{subequations}\label{eqn-2}
	\begin{align}
		\text{(P}&\text{1)}
		\quad
		\max_{\{\alpha_s\}} \quad
		\left(1+\gamma_s\right)\times \left(1+\gamma_w\right),\\
		\notag
		\text{s.}&\text{t.}
		\quad\quad\quad\quad\quad
		\text{(7b), (7c), (7d)};
	\end{align}
\end{subequations}
the objective function of (P1) is denoted as $m(\alpha_s)$, and we proceed to calculate its first derivative as follows
\begin{equation}
	m^{'}(\alpha_s)=
	\tfrac{\rho \cdot (1+\rho \cdot \min{\{\left|h_1(\mathbf{r}_1)\right|^2,\left|h_2(\mathbf{r}_2)\right|^2\}})\cdot \left|\left|h_1(\mathbf{r}_1)\right|^2-\left|h_2(\mathbf{r}_2)\right|^2\right|}
	{\left(\rho \cdot \alpha_s \cdot \min{\{\left|h_1(\mathbf{r}_1)\right|^2,\left|h_2(\mathbf{r}_2)\right|^2\}}+1\right)^2}.
\end{equation}
Additionally, under constraint (7b), we can obtain a lower bound for $\alpha_s$, and two separate upper bounds can be derived based on constraint (7c), as shown bellow.
\begin{align}
	\quad
		\alpha_s & \geqslant \frac{\gamma_0}{\rho \cdot \max{\{\left|h_1(\mathbf{r}_1)\right|^2,\left|h_2(\mathbf{r}_2)\right|^2\}}} \triangleq l_1,\\
		\alpha_s & \leqslant \frac{\rho \cdot \max{\{\left|h_1(\mathbf{r}_1)\right|^2,\left|h_2(\mathbf{r}_2)\right|^2\}}-\gamma_0}
		{\rho \cdot \max{\{\left|h_1(\mathbf{r}_1)\right|^2,\left|h_2(\mathbf{r}_2)\right|^2\}} \cdot (1+\gamma_0)} \triangleq \mu_1,\\
		\alpha_s & \leqslant \frac{\rho \cdot \min{\{\left|h_1(\mathbf{r}_1)\right|^2,\left|h_2(\mathbf{r}_2)\right|^2\}}-\gamma_0}
		{\rho \cdot \min{\{\left|h_1(\mathbf{r}_1)\right|^2,\left|h_2(\mathbf{r}_2)\right|^2\}} \cdot (1+\gamma_0)} \triangleq \mu_2.
\end{align}

Depending on the non-negativity of $m^{'}(\alpha_s)$ and the varied relationships between the lower bound and upper bounds for $\alpha_s$, the optimal solution for (P1) exhibits distinct cases as follows (The user encountering inferior channel conditions is denoted as WU, in contrast to the user experiencing robust channel conditions, designated as SU.).

\emph{Case \uppercase\expandafter{\romannumeral1}}: $l_1\leqslant\mu_2\leqslant\mu_1$, a feasible optimal solution $\alpha_s^{\star}=\mu_2$ for (P1) exists. In cases where the aforementioned relationship cannot be precisely satisfied, an outage event occurs within the communication system.

\emph{Case \uppercase\expandafter{\romannumeral2}}: $0\leqslant\mu_2<l_1\leqslant1$, in this scenario, there must be one user experiencing an outage event. The reference data rate obtained from SU is denoted as $\log_2\left(1+\frac{\rho \cdot \max{\{\left|h_1(\mathbf{r}_1)\right|^2,\left|h_2(\mathbf{r}_2)\right|^2\}}-\gamma_0}{1+\gamma_0}\right)\triangleq R_1$, while for WU, maximal rate is $\log_2 \left(1+\rho \cdot \min{\{\left|h_1(\mathbf{r}_1)\right|^2,\left|h_2(\mathbf{r}_2)\right|^2\}}\right)\triangleq R_2$. The WU encounters an outage event to maximize channel capacity, when $R_1 > R_2$; conversely, the SU experiences an outage event. And we rule that the power allocation coefficient for user in an outage event is reduced to 0.

\emph{Case \uppercase\expandafter{\romannumeral3}}: $\mu_2<0 < l_1 \leqslant 1$, the constraint necessary for accurately decoding signals to the WU remains unsatisfied, resulting in a perpetual outage for the WU. And the optimal $\alpha_s$ which maximizes channel capacity, is denoted as $\alpha_s^{\star}=1$.

\emph{Case \uppercase\expandafter{\romannumeral4}}: $0 \leqslant \mu_2 < 1 < l_1$, similar to Case \uppercase\expandafter{\romannumeral3}, the SU is always in an outage and the optimal $\alpha_s$ is $\alpha_s^{\star} = 0$.

\emph{Case \uppercase\expandafter{\romannumeral5}}: Apart from the aforementioned relationships, both the SU and the WU experience an outage.

\begin{figure*}
	\begin{align}
		&F_i(\mathbf{r}_i)=-\left[\sum_{k=1}^{L_i^r}\mathbf{M}_i[k,k]+\sum_{k=1}^{L_i^r-1}\sum_{l=k+1}^{L_i^r}2\left|\mathbf{M}_i[k,l]\right| \cdot \cos \left(\frac{2\pi}{\lambda} \left(\rho_{i,k}^r(\mathbf{r}_i)-\rho_{i,l}^r(\mathbf{r}_i)\right)-\angle\mathbf{M}_i[k,l]\right)\right],\tag{16}\\
		&\qquad\qquad\qquad\enspace
		F_i(\mathbf{r}_i)\leqslant F_i(\mathbf{r}^n_i) + \triangledown F_i(\mathbf{r}^n_i)^T(\mathbf{r}_i - \mathbf{r}^n_i) + \frac{1}{2}\delta_i(\mathbf{r}_i - \mathbf{r}^n_i)^T(\mathbf{r}_i - \mathbf{r}^n_i),\tag{17}\\
		&\qquad\quad
		\left|\frac{\partial^2F_i}{(\partial x_i^r)^2}\right|
		\leqslant \frac{8\pi^2}{\lambda^2}\sum_{k=1}^{L_i^r-1}\sum_{l=k+1}^{L_i^r}\left|\mathbf{M}_i[k,l]\right| \cdot \left[\cos\theta^r_{i,k}\sin\phi^r_{i,k}-\cos\theta^r_{i,l}\sin\phi^r_{i,l}\right]^2 \triangleq A_i, \tag{18}\\
		&\qquad\qquad\qquad\enspace
		\left|\frac{\partial^2F_i}{(\partial y^r_i)^2}\right|
		\leqslant 	\frac{8\pi^2}{\lambda^2}\sum_{k=1}^{L_i^r-1}\sum_{l=k+1}^{L_i^r}\left|\mathbf{M}_i[k,l]\right| \cdot \left[\sin\theta^r_{i,k}-\sin\theta^r_{i,l}\right]^2 \triangleq B_i, \tag{19}\\
		&\left|\frac{\partial^2F_i}{\partial x_i^r \partial y_i^r}\right|
		\leqslant
		\frac{8\pi^2}{\lambda^2}\sum_{k=1}^{L_i^r-1}\sum_{l=k+1}^{L_i^r}\left|\mathbf{M}_i[k,l]\right| \cdot \left[\cos\theta^r_{i,k}\sin\phi^r_{i,k}-\cos\theta^r_{i,l}\sin\phi^r_{i,l}\right]\cdot \left[\sin\theta^r_{i,k}-\sin\theta^r_{i,l}\right] \triangleq C_i, \tag{20}
	\end{align}
	\hrulefill
\end{figure*}

\subsection{Positions of MAs}
We further consider acquiring the maximal channel capacity by optimizing the positions of MAs for users, when the power allocation coefficient remains constant. The ensuing sub-problem, denoted as (P2), can be succinctly given as
\begin{equation*}\label{eqn-3}
	\begin{aligned}
		\text{(P}&\text{2)}
		\quad\quad
		\max_{\{\mathbf{r}_1, \mathbf{r}_2\}}
		\enspace
		\text{(8a)},\\
		\text{s.}&\text{t.}
		\quad\quad\enspace
		\text{(7b), (7c), (7e)}.
	\end{aligned}
\end{equation*}
Given that The CRC of the SU is influenced by one parameter, either $\mathbf{r}_1$ or $\mathbf{r}_2$, while that of the WU is influenced by the other parameter, (P2) can be decomposed into two sub-problems (P2.1) and (P2.2).
\begin{subequations}\label{eqn-4}
	\begin{align}
		\text{(P2}&\text{.1)}
		\qquad\qquad\qquad
		\max_{\{\mathbf{r}_1,\mathbf{r}_2\}}
		\quad
		1+\gamma_s,\\
		\text{s.}&\text{t.}
		\qquad\qquad\qquad\qquad\quad
		\gamma_s \geqslant \gamma_0,\\
		&\quad\quad\enspace
		\frac{\rho \cdot (1-\alpha_s) \cdot \max{\{\left|h_1(\mathbf{r}_1)\right|^2,\left|h_2(\mathbf{r}_2)\right|^2\}}}
		{\rho \cdot \alpha_s \cdot \max{\{\left|h_1(\mathbf{r}_1)\right|^2,\left|h_2(\mathbf{r}_2)\right|^2\}} +1}\geqslant\gamma_0,\\
		&\quad\quad\quad\quad\quad\quad\quad\quad
		\mathbf{r}_i \in \mathcal{D}_i,\quad i={1,2}.
	\end{align}
\end{subequations}

We observe that the objective function of (P2.1) exhibits monotonic growth concerning $\max{\{\left|h_1(\mathbf{r}_1)\right|^2,\left|h_2(\mathbf{r}_2)\right|^2\}} \triangleq h_s$, and the constraint (13b) imposes a lower bound on $h_s$. Furthermore, it is notable that the left-hand side of the inequality in constraint (13c) also demonstrates monotonic increase with respect to $h_s$, approaching $\frac{1-\alpha_s}{\alpha_s}$ as $h_s$ tends towards infinity. Consequently, in scenarios where $\frac{1-\alpha_s}{\alpha_s} < \gamma_0$, no feasible solutions exist for (P2.1). However, this scenario has already been discussed in Case \uppercase\expandafter{\romannumeral1} to Case \uppercase\expandafter{\romannumeral5}. Therefore, to optimize the objective function of (P2.1), maximizing $h_s$ suffices.

Then, we focus on the sub-problem (P2.2) as following
\begin{subequations}\label{eqn-5}
	\begin{align}
		\text{(P2}&\text{.2)} 
		\qquad\quad
		\max_{\{\mathbf{r}_1, \mathbf{r}_2\}}
		\quad
		1+\gamma_w,
		\\
		\text{s.}&\text{t.} \qquad\qquad\quad\enspace
		\gamma_w \geqslant \gamma_0,\\
		\quad&
		\qquad\qquad
		\mathbf{r}_i \in \mathcal{D}_i,
		\quad i={1, 2};
	\end{align}
\end{subequations}
similar to (P2.1), to maximize the objective of (P2.2), our focus lies in maximizing $\min{\{\left|h_1(\mathbf{r}_1)\right|^2,\left|h_2(\mathbf{r}_2)\right|^2\}}$. Consequently, to tackle sub-problems (P2.1) and (P2.2), it suffices to deal with two sub-problems possessing identical structures, denoted as (P3.1) and (P3.2), as illustrated bellow.
\begin{subequations}
	\begin{align}
		\text{(P3}&\text{.}i\text{)}\quad\quad
		\min_{\{\mathbf{r}_i\}}\quad
		F_i(\mathbf{r}_i)\triangleq-\left|h_i(\mathbf{r}_i)\right|^2,\\
		\text{s.}&\text{t.}\quad\qquad\quad\qquad
		\mathbf{r}_i \in \mathcal{D}_i.
	\end{align}
\end{subequations}

According to the equation (4), the objective function of (P3.$i$) can be transformed into equation (16), where we define $\mathbf{M}_i=\mathbf{\Sigma}_i\cdot\mathbf{G}_i\cdot\mathbf{1}\cdot\mathbf{1}^T\cdot\mathbf{G}_i^H\cdot\mathbf{\Sigma}_i^H$.

To address the non-convex nature of the problem (P3.i), we adopt a method based on SCA technique to optimize the positions of MAs for users. Drawing on Taylor's theorem, we construct a global upper bound of $F_i$ given the position $\mathbf{r}_i^n$ of the i-th user's MA in the n-th iteration. The bound as shown in (18) is ensured by introducing a positive real number $\delta_i$ satisfying $\delta_i\mathbf{I}_2 \succeq \triangledown^2F_i(\mathbf{r}_i)$. Based on $\Vert \triangledown^2F_i(\mathbf{r}_i) \Vert^2_2 \leqslant \Vert \triangledown^2 F_i(\mathbf{r}_i) \Vert^2_F$ and derivations (18)-(20), we can ascertain the value of $\delta_i$ as $\delta_i = (A_i^2 + B_i^2 + 2C_i^2)^{\frac{1}{2}}$.

Therefore, problem (P3.$i$) is reduced to the problem (P4.$i$) as
\begin{subequations}
	\begin{align}\tag{21}
		\text{(P4}&\text{.}i\text{)} \quad
		\min_{\{\mathbf{r}_i\}} \enspace
		F_i(\mathbf{r}^n_i) + \triangledown F_i(\mathbf{r}^n_i)^T(\mathbf{r}_i - \mathbf{r}^n_i) \\
		\notag
		 & \quad\quad\quad\quad\quad
		 + \frac{1}{2}\delta_i(\mathbf{r}_i - \mathbf{r}^n_i)^T(\mathbf{r}_i - \mathbf{r}^n_i),\\
		 \notag
		 \text{s.}&\text{t.}
		 \quad\qquad\qquad\qquad\quad
		 (16b).
	\end{align}
\end{subequations}
Ultimately, we can transform the optimization of non-convex problems (P3.1) and (P3.2) into optimizing a series of convex problems (P4.1) and (P4.2), wherein the optimal solution can be readily attained.

\subsection{Algorithm}
The algorithm for jointly optimizing the power allocation coefficient and positions of MAs is delineated in \textbf{Algorithm 1}. Initially, we derive the optimal CRCs through the optimization of MA positions. Subsequently, the allocation coefficient is determined by identifying the occurrence of an outage event under different scenarios. Let $\vartheta_1$ and $\vartheta_2$ denote the number of SCA iterarions for user1 and user2 respectively. The total complexity of \textbf{Algorithm 1} is $\mathcal{O}(\vartheta_1(L_1^r)^2+\vartheta_2(L_2^r)^2+NL_1^tL_1^r+NL_2^tL_2^r)$.
\begin{algorithm}
	\caption{Joint power allocation and antenna position design}
	Initial the positions of MA, $\mathbf{r}_1^0, \mathbf{r}_2^0$; choose a stepsize $\eta \in (0, 1)$, a criteria $\epsilon$ and a maximal iteration $\vartheta$; and set $n=0$\\
	1: \textbf{repeat}\\
	2: \quad Set $n\longleftarrow n+1$\\
	3: \quad Set $\hat{\mathbf{r}}^n_i$ to be an optimal solution of P4.i\\
	4: \quad Set $\mathbf{r}_i^n = \eta\hat{\mathbf{r}}_i^n+(1-\eta)\mathbf{r}_i^{n-1}$\\
	5: \textbf{until} $\Vert \mathbf{r}_i^n - \mathbf{r}_i^{n-1}\Vert_2\leqslant\epsilon$ or $n \geqslant \vartheta$\\
	6: Set $\mathbf{r}_i = \mathbf{r}^n_i$ and calculate $l_1$, $\mu_1$, $\mu_2$\\
	7: Determine the case for P1 and allocation coefficient $\alpha_s$
\end{algorithm}

\section{Simulation Results}

\begin{figure*}
	\centering
	\begin{minipage}{0.329\linewidth}
		\centering
		\includegraphics[width=1\linewidth]{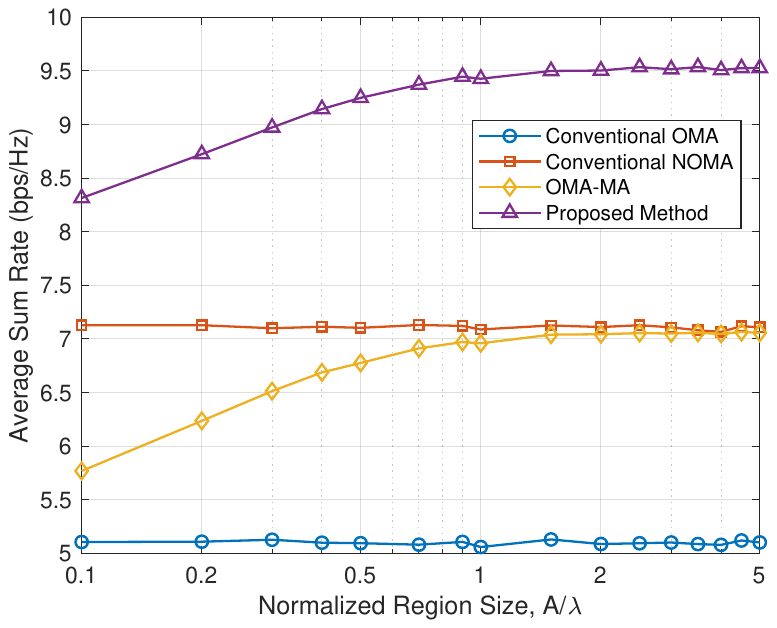}
		\caption{Average sum rate versus normalized region size.}
	\end{minipage}
	\begin{minipage}{0.329\linewidth}
		\centering
		\includegraphics[width=1\linewidth]{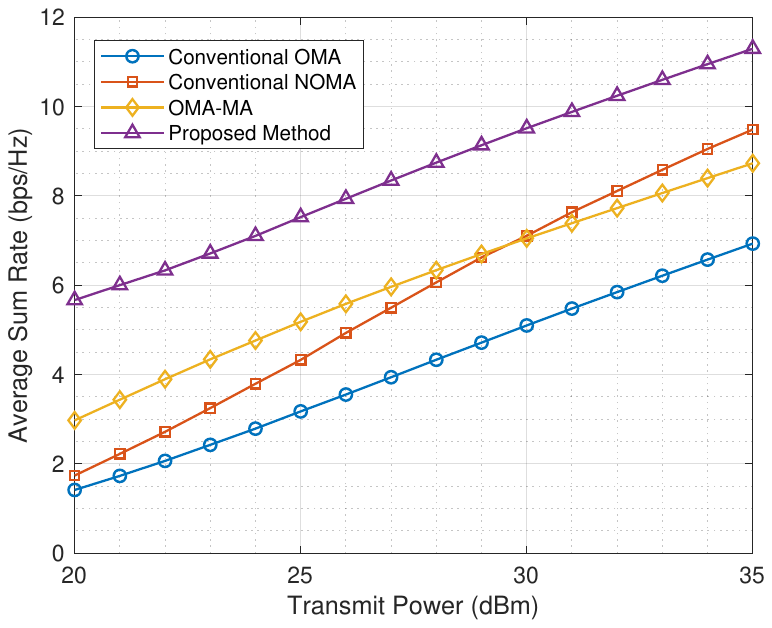}
		\caption{Average sum rate versus transmit power.}
	\end{minipage}
	\begin{minipage}{0.329\linewidth}
		\centering
		\includegraphics[width=1\linewidth]{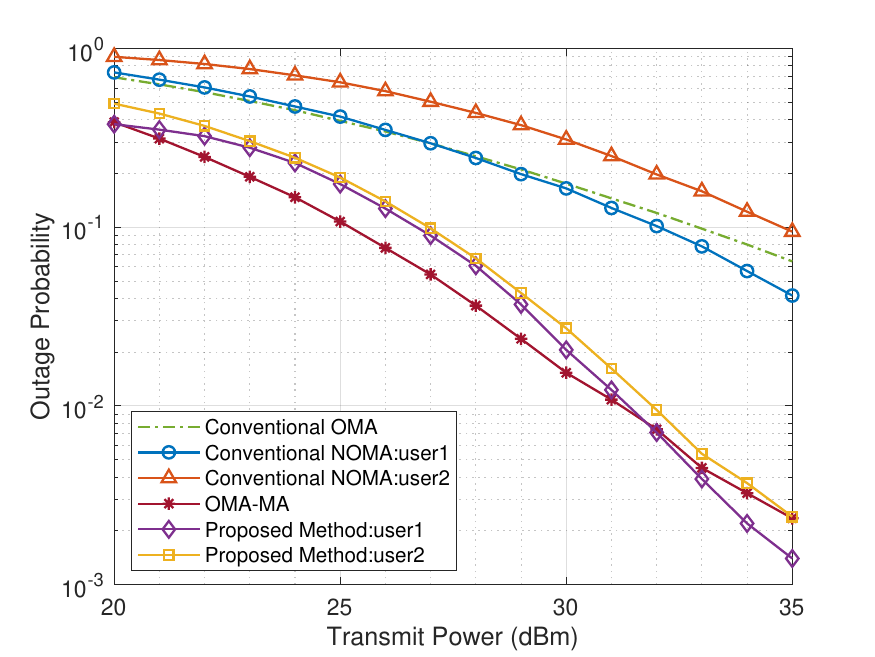}
		\caption{Outage probability versus transmit power.}
	\end{minipage}
\end{figure*}

In this section, we validate the proposed design through a series of numerical simulations. Specifically, we consider a communication system wherein a BS with an FPA array consisting of $N=16$ elements, catering to 2 users. The distances from user1 and user2 to the BS are denoted as $d_1=60$ m and $d_2=100$ m respectively. The carrier frequency utilized from the BS to users is 3 GHz, corresponding to a carrier wavelength of $\lambda=0.1$ m. We assume that the number of transmit paths equals that of receive paths, denoted as $L_i^t=L_i^r=L=10$. The PRM for each user is a diagonal matrix, with diagonal elements being mutually independent and conforming to the same distribution $\mathcal{CN}(0,c_0d^{-\beta}L^{-1})$, where $c_0=(\frac{\lambda}{4\pi})^2$ denotes the expected value of average channel gain at the reference distance of $1$ m and $\beta=2.8$ signifies the path-loss factor.  The power of AWGN is $\sigma^2=-90$ dBm. The elevations and azimuths of AoAs/AoDs are presumed to be independent and identically distributed, adhering to a uniform distribution over the interval $[-\frac{\pi}{2},\frac{\pi}{2}]$. Moreover, the SINR threshold is $\gamma_0=10$ dB.

We consider three baseline schemes for comparison: (1) conventional OMA, where each user utilizes an FPA and OMA technique; (2) conventional NOMA, where each user employs an FPA and NOMA technique; and (3) OMA-MA, where each user incorporates an MA and utilizes OMA technique.

Fig. 2 illustrates the achievable average sum rate for each scheme in relation to the normalized region size. With the BS's transmit power set at $P=30$ dBm, it is evident that the average sum rates increase alongside the normalized region size for both our proposed method and OMA-MA. Additionally, it is notable that the average sum rate for these two schemes tend to stabilize once the normalized region size reaches 1. This observation provides a useful reference point for setting the region size. It is noteworthy that the average sum rate tends to converge once the normalized region size reaches 1 for both conventional NOMA and OMA-MA. Furthermore, it is found that the sum rates achieved by conventional OMA and NOMA schemes remain unchanged for different region sizes, as the fixed antennas at the receivers in these two schemes cannot utilize the new spatial DoFs. Among all schemes considered, our proposed method demonstrates the highest achievable average sum rate within the same normalized region size.

Fig. 3 shows the achievable sum rate for each scheme versus the transmit power of the BS. When considering a set of $A=3\lambda$, it can be seen that the achievable average sum rates increase across all schemes with the rise in transmit power. This augmentation can be ascribed to the heightened SINR achieved by increasing the transmit power. Notably, the average sum rate for conventional NOMA falls below that of OMA-MA at low transmit power, but the trend reverses at high transmit power levels. This observation suggests that the performance of conventional NOMA is inferior to that of OMA with MA due to the high outage probability of NOMA. However, the potential of NOMA adoption becomes more apparent when transmit power is ample. Similarly, our proposed method exhibits the highest average sum rate among all schemes at equivalent transmit power levels.

Finally, Fig. 4 exhibits the variation in outage probabilities for user1 and user2 concerning transmit power under the setup where $A=3\lambda$. It can be observed that the outage probability in schemes adopting NOMA is relatively higher than in those adopting OMA when the transmit power is low. As the transmit power increases, the outage probability in all schemes decreases. However, when the transmit power is high, the outage probability for user1 in schemes using conventional NOMA and the proposed method is lower than in schemes using conventional OMA and OMA-MA, respectively. This phenomenon occurs because NOMA causes the signal sent to SU to interfere when decoding the signal from WU. To ensure the communication quality of WU, more power is allocated to WU, resulting in less power allocated to SU compared to OMA. Therefore, at low transmission power, the interruption probability for both users is higher than that in OMA. As the power increases, it provides more redundancy for the implementation of NOMA, highlighting its advantages. Moreover, our power allocation strategy results in the outage probabilities of user1 and user2 being consistently closer to each other compared to conventional NOMA, particularly at low transmission power levels.

\section{Conclusion}
In this paper, we considered an MA-assisted NOMA downlink communication system wherein a BS is equipped with FPAs and users possess a single MA.  We focused on maximization of the channel capacity by joint optimization of power allocation and antenna positions. We presented an efficient algorithm leveraging the principle of AO and SCA to tackle this non-concave problem.  Simulation results illustrated our proposed design can achieve better channel sum rate and lower outage probability than that of conventional systems. In future work, a significant open research question is how to ensure power allocation fairness and optimal arrangement of multiple users in MA-assisted NOMA schemes.

\end{document}